\documentclass[ejp]{webofc}
\usepackage[varg]{txfonts}   
\usepackage{color}
\woctitle{INPC 13}
\begin{document}
\title{Cause of the charge radius isotope shift at the \emph{N}=126 shell gap}
\author{P.M. Goddard \and P.D. Stevenson \and A. Rios
}                   \institute{Department of Physics, University of Surrey, Guildford, GU2 7XH, UK}

\abstract{
We discuss the mechanism causing the `kink' in the charge radius isotope shift at the $N=126$ shell closure. The occupation of the 1$i_{11/2}$ neutron orbital is the decisive factor for reproducing the experimentally observed kink.  We investigate whether this orbital is occupied or not by different Skyrme effective interactions as neutrons are added above the shell closure. Our results demonstrate that several factors can cause an appreciable occupation of the 1$i_{11/2}$ neutron orbital, including the magnitude of the spin-orbit field, and the isoscalar effective mass of the Skyrme interaction. The symmetry energy of the effective interaction has little influence upon its ability to reproduce the kink.
}
\maketitle
\section{Introduction}
\label{intro}
The kink in the charge radius shift of even-even nuclei at the N=126 shell closure has been challenging to explain theoretically for several decades. Popular explanations advocate the influence of a neutron $2g_{9/2}$ level with a large orbital radius \cite{Reinhard,Sharma}, or the role of ground-state quadrupole correlations \cite{Brownii}.
\begin{figure}[h!]
\begin{centering}
\includegraphics[width=12.5cm]{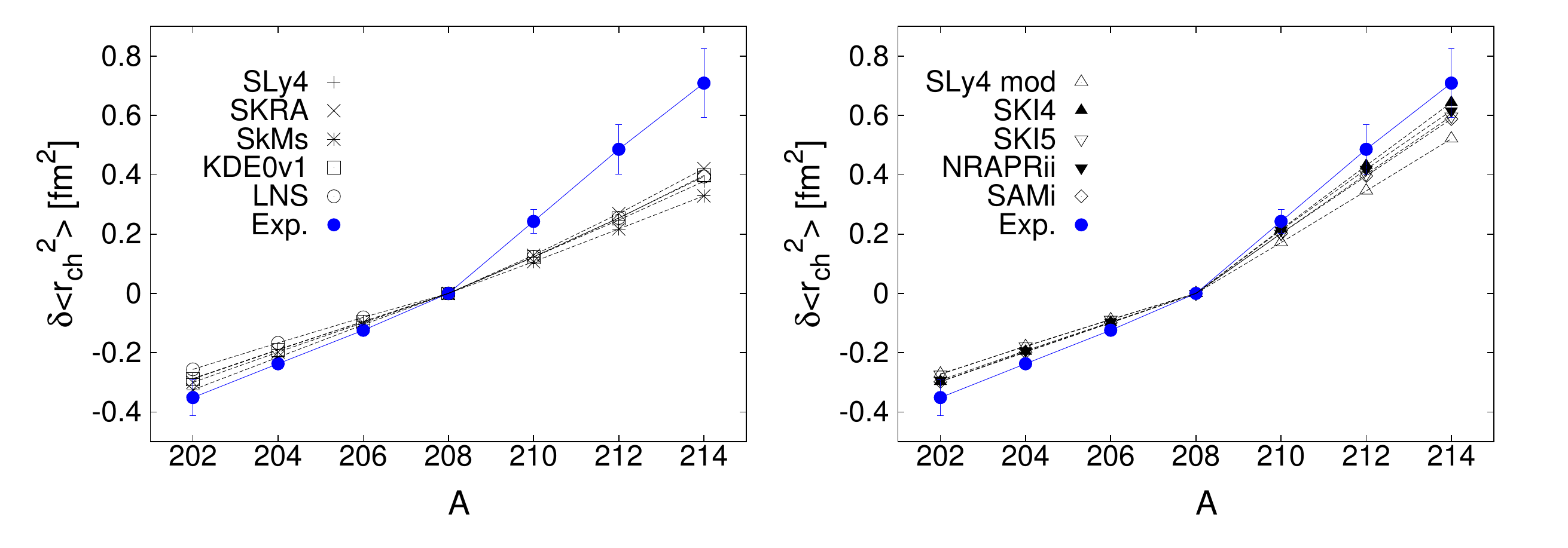}
\caption{Charge radius isotope shift for lead isotopes using various Skyrme forces, in comparison to experimental data (circles). The reference nucleus is $^{208}$Pb. Unsuccessful Skyrme interactions are shown in the left panel, successful ones in the right panel.}
\label{fig:kink}     
\end{centering}
\end{figure}

We define the charge radius isotope shift {$\delta  \langle r^2_{\textrm{ch}} \rangle$} as the differences between the mean squared charge radius, $\langle r^2_{\textrm{ch}} \rangle$, of a series of isotopes and that of a given reference isotope, $^{208}$Pb in this case. In Fig.~\ref{fig:kink}, the charge radius isotope shift in the Pb chain for several Skyrme effective interactions is shown in comparison to experimental data \cite{angeli}.
Most non-relativistic mean-field models struggle to reproduce the kink \cite{Reinhard}. 
In Ref.~\cite{God13} a new explanation is presented, advocating the influence of the 1$i_{11/2}$ neutron orbital upon tightly bound $n=1$ proton orbitals due to the spatial overlap. Skyrme interactions which result in a significant population of the 1$i_{11/2}$ orbital for isotopes past the $N =126$ shell closure reproduce the kink. However, there are many factors within the Skyrme interaction that can cause this orbital to be preferentially occupied. We will discuss this briefly herein.

Experimentally, the gradient of the linear fit to the isotope shift data approximately doubles after the $N=126$ shell closure \cite{angeli}. The ratio R$_{kink}$ of the gradient above $A=208$ compared to below $A=208$ is $2.04(5)$. We define `successful' Skyrme interactions in this context as those that produce a ratio R$_{kink}$ within $\sim 10 \, \%$ of this value.

\section{Occupation of the neutron $1i_{11/2}$ orbital}
\label{sec:1}
The occupation of the 1$i_{11/2}$ orbital is the decisive factor in producing the kink. In our calculations, pairing is determined using the BCS method, which provides non-zero occupation of orbitals slightly above the Fermi surface beyond $A=208$.  Above the $N=126$ shell closure, the next two most bound neutron orbitals are typically the $2g_{9/2}$ and the $1i_{11/2}$. Experimentally, the $2g_{9/2}$ orbital is more deeply bound than the $1i_{11/2}$ orbital \cite{brown}, though separated only by approximately $1$ MeV. In general, however, mean-field single-particles energies need not necessarily agree with measured energy levels \cite{Duguet2012}. Within the BCS framework, one expects that whichever of the two orbitals has the greatest binding energy will have the greatest occupation probability.  

\begin{figure}[h!]
\begin{centering}
\includegraphics[width=12.5cm]{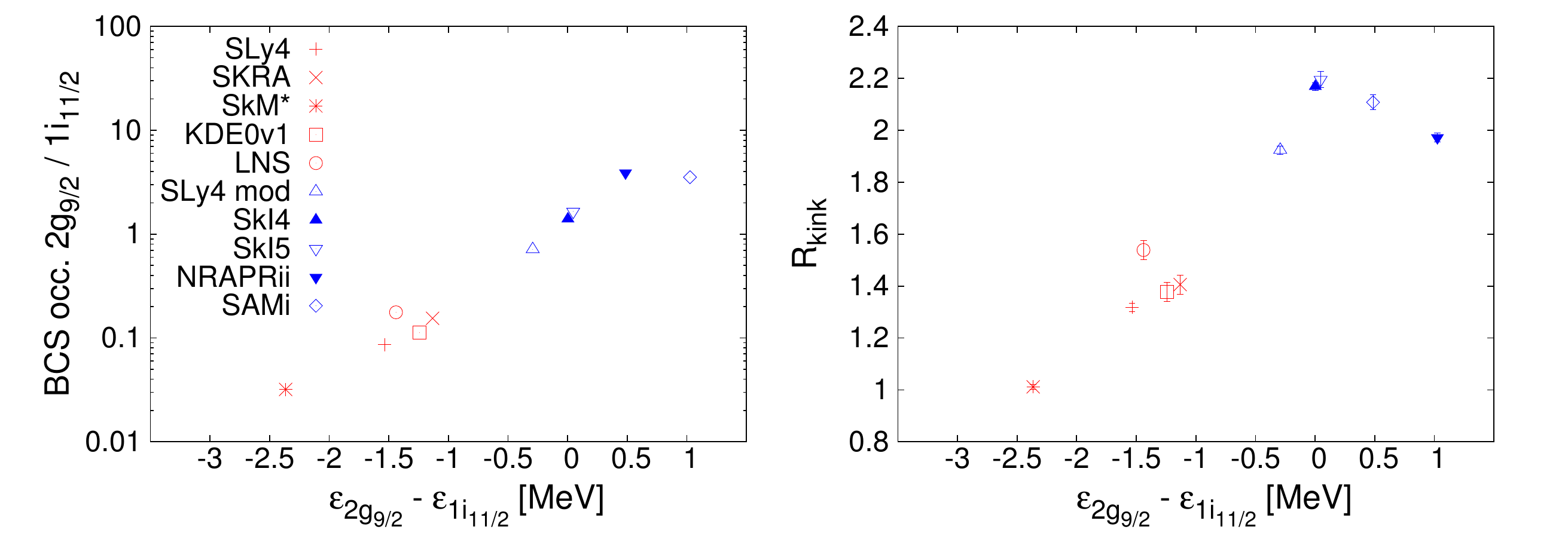}
\caption{Left panel: ratio of BCS occupation probability of the neutron $2g_{9/2}$ level compared to the 1$i_{11/2}$ in $^{210}$Pb as a function of the energy difference of the orbitals, $\epsilon_{2g_{9/2}} - \epsilon_{1i_{11/2}}$ for various Skyrme interactions. Right panel: R$_{kink}$ as a function of the energy difference. A a positive energy difference corresponds to the $1i_{11/2}$ being more bound than the $2g_{9/2}$. Skyrme interactions that reproduce the kink are shown in blue, and those that fail in red.}
\label{fig:edif}     
\end{centering}
\end{figure}

Indeed, we show in the left panel of Fig.~\ref{fig:edif} the ratio of the BCS calculated occupation probability of the $2g_{9/2}$ compared to the $1i_{11/2}$ orbitals in  the case of $^{210}$Pb. The plot suggest an exponential dependence of the occupation as a function of single-particle energy. The right panel concentrates on R$_{kink}$. A linear relation is observed between the ratio of slopes and the single-particle energy difference. Skyrme interactions which reproduce the kink always have a significant BCS occupation of the $1i_{11/2}$ neutron orbital compared to the $2g_{9/2}$. Within the mean-field framework, the only way this can occur is when the $1i_{11/2}$ orbital is either more bound than or virtually degenerate with the $2g_{9/2}$. Factors which influence the energy level ordering are therefore crucial for correct reproduction of the kink. We shall discuss some of these factors in the following subsections.

\subsection{Spin-orbit field}
\label{sec:1.1}
Several authors have identified the importance of the spin-orbit field for reproducing the kink from comparisons between relativistic and non-relativistic models \cite{Reinhard,Sharma}. Their interpretation of the mechanism resulting in the kink, however, differs from that in Ref.~\cite{God13}. The old argument revolves around occupation of the $2g_{9/2}$ neutron orbital being the decisive factor. Skyrme interactions with a weak spin-orbit field will cause this state to be less bound, and therefore have a larger radius. The resulting influence on the proton states via the symmetry energy will be more pronounced when the radius is larger, and might be able to explain the sudden change in charge radius past the $N=126$ shell closure.

We interpret the influence of the spin-orbit field differently. A way to ensure the neutron $1i_{11/2}$ orbital is more bound than the $2g_{9/2}$ is by reducing the magnitude of the spin-orbit field. With a small enough spin-orbit contribution, the spitting of the $2g_{9/2}$ and $2g_{7/2}$ will be reduced enough to ensure that the 1$i_{11/2}$ will be more bound than the $2g_{9/2}$. As shown in Fig.~\ref{fig:edif}, the more tightly bound state will have the higher occupation probability within the mean-field model.
\begin{figure}[h!]
\begin{centering}
\includegraphics[width=12.5cm]{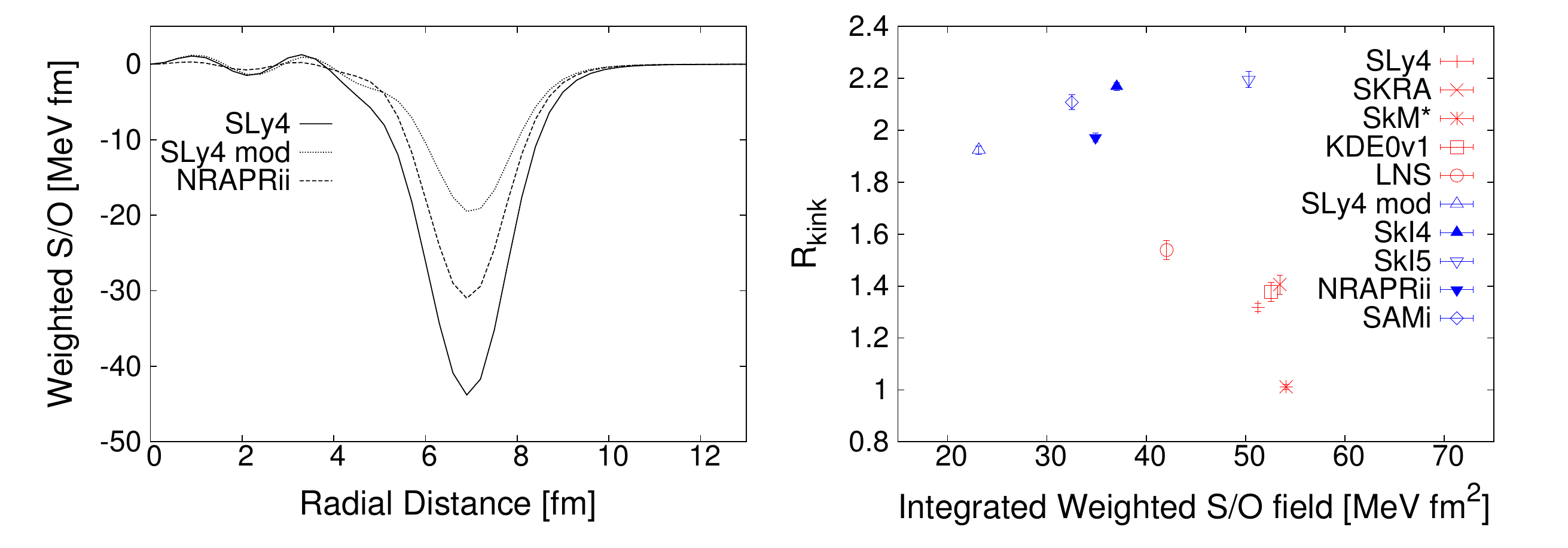}
\caption{Left panel: typical radial spin-orbit fields (weighted by a factor of r) for a selection of Skyrme interactions. Right panel: R$_{kink}$ as a function of the integrated weighted spin-orbit field.}
\label{fig:wo}     
\end{centering}
\end{figure}

Shown in the left panel of Fig.~\ref{fig:wo} are the radial spin-orbit fields for a selection of Skyrme interactions in $^{210}$Pb. 
Other than the magnitude, there are no major differences between the radial dependence of spin-orbit fields whether they reproduce the kink or not. The relation between the integrated spin-orbit fields and $R_{kink}$ is displayed in the right panel. Typically, those Skyrme interactions with a smaller integrated spin-orbit field reproduce the kink. The main exception to this is SkI5, which manages to reproduce the kink even with a spin-orbit field of a similar magnitude to that of unsuccessful interactions. This demonstrates that the spin-orbit field is not necessarily the only factor contributing to the success of a Skyrme interaction in reproducing the kink.

\subsection{Nuclear matter properties at saturation density}
\label{sec:1.2}
\begin{figure}[h!]
\begin{centering}
\includegraphics[width=12.5cm]{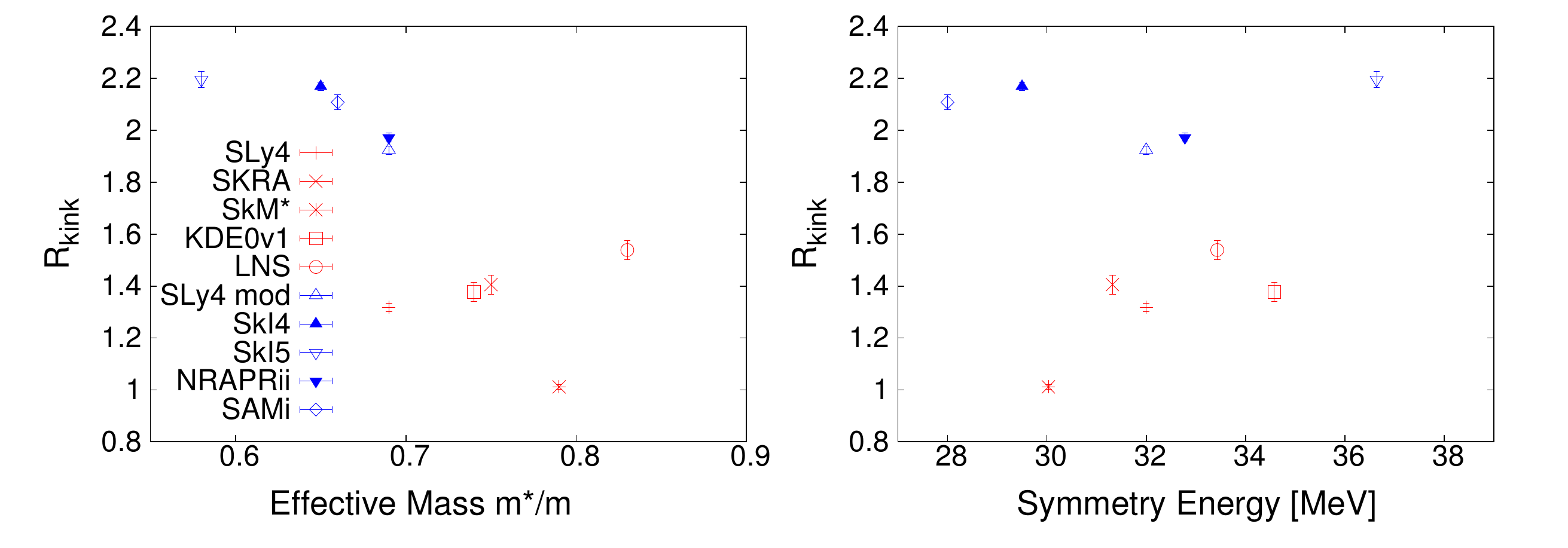}
\caption{Left panel: R$_{kink}$ as a function of isoscalar effective mass, $m^*/m$.  Right panel: R$_{kink}$ as a function of the symmetry energy, $J$.}
\label{fig:nm}  
\end{centering}
\end{figure}

Isovector nuclear matter properties may be influential upon successfully reproducing the kink. For brevity, we will discuss only two: the effective mass and the symmetry energy. The symmetry energy is related to the influence the relevant occupied neutron orbital past the $N=126$ closure will have upon the deeply-bound proton orbitals. The effective mass relates to the density of the energy levels around the Fermi surface \cite{Dutra}. The ratio $R_{kink}$ is shown as a function of these two parameters in Fig.~\ref{fig:nm}. 

Ref.~\cite{Reinhard} mentions that the interaction SkI5 may be successful in reproducing the kink due to its unrealistically small effective mass. The small effective mass leads to virtually degenerate $1i_{11/2}$ and $2g_{9/2}$ orbitals, which leads to both having an appreciable occupation probability. Therefore, our explanation of the kink being reproduced due to occupation of the $1i_{11/2}$ applies. 
The right panel of Fig.~\ref{fig:nm} displays no correlation between the symmetry energy and the ability of the Skyrme interaction to reproduce the kink. This demonstrates that even for Skyrme interactions with a smaller symmetry energy, all that is required to reproduce the kink is an appreciable occupation of the $1i_{11/2}$.

\section{Conclusions}
\label{sec:2}

We have presented further evidence that the occupation of the $1i_{11/2}$ is the pivotal factor in the reproduction of the well-known kink in the isotopic shift of the charge radius around the $N=126$ shell closure. Further discussion has been given with regards of the mechanism resulting in a significant occupation of this orbital. In particular, we have looked at correlations between the kink strength and the spin-orbit field as well as the symmetry energy and effective mass. There may be other factors which can result in the occupation of this orbital. For example, inclusion of the tensor force is known to have significant effect upon the spin-orbit splitting \cite{emma}. Beyond mean-field considerations are also likely to have an important effect upon the occupation of levels past the $N=126$ closure.


\begin{thebibliography}{}
%
%
\bibitem{Reinhard}
Reinhard, P.-G. and Flocard, H., Nucl. Phys. \textbf{A584}, 467 (1995)
\bibitem{Sharma}
Sharma, M. M. and Lalazissis, G. and K\"onig, J. and Ring, P., Phys. Rev. Lett. \textbf{74}, 3744 (1999)
\bibitem{Brownii}
Brown, B. A. and Bronk C. R. and Hodgson P.E., J. Phys. G \textbf{10}, 1683 (1984) 
\bibitem{angeli}
Angeli, I., At. Data Nucl. Data Tables \textbf{87}, 185 - 206 (2004)
\bibitem{God13}
Goddard, P. M. and Stevenson, P. D. and Rios, A., Phys. Rev. Lett. \textbf{110}, 032503 (2013)
\bibitem{Duguet2012}
Duguet, T. and Hagen, G., Phys. Rev. C \textbf{85}, 034330 (2012)
\bibitem{brown}
Brown, B. A., Phys. Rev. C \textbf{58}, 220--231 (1998)
\bibitem{Dutra}
Dutra, M. et al., Phys. Rev. C \textbf{85}, 035201 (2012)
\bibitem{emma}
Suckling, E. B. and Stevenson, P. D., EPL \textbf{90}, 12001 (2010)
\end{thebibliography}
\end{document}